\documentclass[12pt]{article}
\usepackage{amsmath}
\usepackage{graphicx}
\usepackage{enumerate}
\usepackage{tabularx}
\usepackage[numbers,square,sort&compress]{natbib}
\usepackage{url} 

\newcommand{\E}{\mathbbm{E}}     

\usepackage{xcolor}

\usepackage{amsthm, amsmath, amssymb, bbm, booktabs, graphicx, tikz}
\usepackage{multirow}
\usepackage{float}
\usepackage{diagbox}

\newtheorem{lemma}{Lemma}
\newtheorem{theorem}{Theorem}

\usepackage{xr-hyper} 
\usepackage{hyperref}  

\usepackage{pgf,tikz}
\usetikzlibrary{
  arrows,
  arrows.meta,
  calc,
  decorations.pathmorphing,
  decorations.pathreplacing,
  fit,
  positioning,
  shapes,
  shapes.arrows,
  shapes.geometric,
  shapes.multipart,
  swigs
}

\pgfkeys{/tikz/swig vsplit/.code={%
  \pgfkeys{/tikz/swig vsplit/.cd,#1}%
}}

\pgfkeys{/tikz/swig vsplit/line color left/.initial = black}
\pgfkeys{/tikz/swig vsplit/line color right/.initial = black}
\pgfkeys{/tikz/swig vsplit/fill color left/.initial = white}
\pgfkeys{/tikz/swig vsplit/fill color right/.initial = white}
\pgfkeys{/tikz/swig vsplit/gap/.initial = 0.8em}
\pgfkeys{/tikz/swig vsplit/line width left/.initial = \the\pgflinewidth}
\pgfkeys{/tikz/swig vsplit/line width right/.initial = \the\pgflinewidth}
\pgfkeys{/tikz/swig vsplit/inner line width left/.initial = 0pt}
\pgfkeys{/tikz/swig vsplit/inner line width right/.initial = 0pt}

\usepackage{scalefnt}

\tikzset{
  -Latex,
  auto,
  node distance=1cm and 1cm,
  semithick,
  state/.style={circle, draw, minimum width=0.5cm},
  point/.style={circle, draw, inner sep=0.04cm, fill, node contents={}},
  bidirected/.style={Latex-Latex, black},
  el/.style={inner sep=2pt, align=left, sloped}
}
\usepackage{bbm}
\newtheorem{assm}{Assumption}

\begin{document}

\def\spacingset#1{\renewcommand{\baselinestretch}%
{#1}\small\normalsize} \spacingset{1}


\title{\bf Designing a quasi-experiment to study the clinical impact of adaptive risk prediction models}

  \author{
    Valerie Odeh-Couvertier \\
    Department of Industrial \& Systems Engineering \\
    University of Wisconsin Madison \\
    \\
    Gabriel Zayas-Cab\'{a}n \\
    Department of Industrial \& Systems Engineering \\
    BerbeeWalsh Department of Emergency Medicine \\
    University of Wisconsin Madison \\
    \\  
    Brian Patterson \\
    BerbeeWalsh Department of Emergency Medicine \\
    University of Wisconsin Madison \\
    \\
    Amy L Cochran\\
    Department of Mathematics \\
    Department of Population Health Sciences\\
    University of Wisconsin Madison
    }
  \maketitle

\bigskip

\noindent%
\vfill

\newpage

\begin{abstract}
Clinical risk prediction is a valuable tool for guiding healthcare interventions toward those most likely to benefit. Yet, evaluating the pairing of a risk prediction model with an intervention using randomized controlled trials presents substantial challenges, making quasi-experimental designs an attractive alternatives. Existing designs, however, assume that both the model and the decision rules used to trigger interventions (typically a risk threshold) remain fixed. This limits their utility in modern healthcare, where both are routinely updated. We introduce a regression discontinuity framework that accommodates adaptation in both the model and the risk threshold. We precisely characterize the form of interference introduced by these adaptations and exploit this structure to establish conditions for identification and thus design estimation strategies. The key idea is to define counterfactual risks---the scores patients would have received under hypothetical reorderings---thereby restoring local exchangeability and enabling valid estimation of the local average treatment effect. Our estimator leverages the fact that, although counterfactual risk vectors grow with time, they typically lie in a low-dimensional space. In simulations of cardiovascular prevention programs, we show that our method accurately recovers treatment effects even as thresholds adapt to meet operational or clinical targets and models are updated to align predicted and observed outcomes or to exclude demographic predictors such as race.
\end{abstract}

\section*{Introduction}

As healthcare systems rapidly adopt artificial intelligence and machine learning tools, evidence of their effectiveness has lagged. Nowhere is this gap more apparent than with clinical risk prediction models. Despite a nearly fourfold rise in publications over three decades, only a small fraction evaluate the impact of these models on patient outcomes~\cite{challener2019proliferation,atkins2022developing}. Most models are implemented based solely on predictive accuracy, without rigorous prospective evaluation of their effectiveness and safety \cite{greene2012implementing,friedman2010achieving,etheredge2007rapid,kelly2019key}. Over 90\% of FDA-approved AI/ML tools lack prospective surveillance, and few report detailed performance data across diverse patient populations~\cite{muralidharan_scoping_2024}. With governing bodies currently drafting standards for clinical AI/ML use~\cite{moons2015transparent,collins2015transparent,reilly2006translating,wolff2019probast}, this is an opportune moment to shape how these tools are evaluated.

To meaningfully assess whether a risk model improves clinical outcomes, we first specify how its predictions are meant to alter clinical care~\cite{atkins2022developing}. As emphasized in both trial design and causal inference, this requires a well-defined intervention \cite{rubin1980randomization,rubin1986statistics}. 
Many risk models illustrate the importance of this connection \cite{motzer1999survival,mekhail2005validation,frank2002outcome,leibovich2003prediction,o2018international,o2014novel,saaristo2005cross,lip2010refining,kanis2004meta,kanis2007use,kanis2008frax,mccarthy2015predictive,motzer2010phase,rini2019atezolizumab,ljungberg2015eau,maron2019enhanced,bartoli2011oral,hindricks20212020,cosman2014clinician,reger2019integrating}: the CHA$_2$DS$_2$-VASc score estimates stroke risk in patients with atrial fibrillation to guide anticoagulation~\cite{lip2010refining,hindricks20212020}; the Fracture Risk Assessment Tool estimates 10-year fracture risk to inform osteoporosis treatment~\cite{kanis2004meta,kanis2007use,kanis2008frax,cosman2014clinician}; and the Veterans Health Administration identifies patients at high risk for suicide using a risk model that triggers safety planning interventions~\cite{mccarthy2015predictive,reger2019integrating}.

Yet even when a model is paired with a specific clinical action, that relationship tends to evolve. Prediction models degrade in accuracy when applied outside the population, setting, or time for which they were developed \cite{van2023there}. Declines in accuracy stem from changing data quality, evolving clinical practices, and new health threats like Covid-19 \cite{binuya2022methodological,nakatsugawa2019needs,van2023there,subbaswamy2020development}. These issues are compounded by historic underrepresentation of women and minorities in clinical datasets \cite{dresser1992wanted, meltzer_what_2008}, resulting in reduced accuracy for these groups. To maintain relevance, models must adapt to the diverse populations and evolving conditions they encounter, using well-established strategies such as recalibration or revision
\cite{steyerberg2004validation,janssen2008updating,binuya2022methodological,binuya2022methodological, su2018review}.

Calls are mounting for research designs capable of evaluating risk models paired with interventions in dynamic clinical environments \cite{atkins2022developing, li2020developing, magrabi2019artificial, horwitz2019creating, heys2023development}. Although randomized controlled trials (RCTs) remain the gold standard, they are often too costly, time-consuming, and rigid to keep pace with the rapid iteration required in this space \cite{horwitz2019creating, atkins2022developing}. Quasi-experimental designs offer a more nimble solution. Among these, regression discontinuity (RD) designs are particularly compelling \cite{thistlethwaite1960regression}. When treatment decisions depend strictly on crossing a specific risk threshold, comparing patients immediately above and below this point allows estimation of local causal effects nearly as reliable as an RCT \cite{hahn2001identification}. RD designs have already been explored for evaluating statin prescribing based on cardiovascular risk and post-discharge interventions driven by readmission risk scores \cite{robinson2015evaluation, robinson2017observational}.

However, traditional RD designs assume \textit{no interference}: that one patient's treatment assignment does not influence another patient's potential outcomes~\cite{rubin1980randomization,rubin1986statistics}. This assumption is violated when models update over time in response to prior patient data, causing subsequent predictions and treatments to become interdependent. Existing RD approaches that handle interference either forgo point estimates or target different causal estimands rather than the local average treatment effect~\cite{rosenbaum2007interference,cattaneo2016inference,aronow2017regression}. Overcoming this critical methodological gap is central to our proposed approach.

In this paper, we introduce an RD design specifically developed to evaluate risk prediction models paired with well-defined interventions that continually learn and adapt. We establish formal identification conditions under interference and propose estimators that robustly recover local causal effects. We conduct simulations centered on a cardiovascular risk prediction model \cite{goff20142013}, exploring a range of plausible scenarios that health systems may encounter when using such models to guide clinical interventions. Finally, informed by these simulations, we provide concrete guidance for researchers and practitioners aiming to evaluate adaptive prediction-based interventions in real-world clinical environments.

\section*{Results}

\subsection*{An adaptive RD design} Our research design involves patients arriving sequentially to a healthcare unit, such as a primary care clinic. Each patient is assessed using a risk prediction model that maps their covariates (e.g., age, symptoms) to a continuous risk score. The model is triggered by a specific event, such as entering the admission decision into the electronic health record (EHR). If the predicted risk exceeds a prespecified threshold, a clinical action is taken. An outcome is then observed at a later time. This outcome need not match the event being predicted by the model. This flexibility is important when long-term risks are predicted, but only short-term outcomes are feasible. Both the model and its threshold may adapt over time.

Table~\ref{tab:rd_emulation} summarizes the design's key elements, illustrated using a cardiovascular risk prediction example. The structure mirrors that of a clinical trial, including common elements like inclusion/exclusion criteria, intervention, comparator, allocation, follow-up, and outcomes. It also incorporates elements specific to our adaptive RD design: the risk prediction model, the trigger event for applying it, and procedures for updating both the model and threshold. In the example, adults aged 40--79 without prior cardiovascular disease are included. At each annual primary care visit (trigger event), a risk model (e.g., pooled cohort equations) estimates 10-year risk of atherosclerotic cardiovascular disease (ASCVD) \cite{goff20142013}. If risk exceeds a threshold, the patient is automatically referred to a preventive care program (intervention); otherwise, routine care continues (comparator). The threshold may be updated to target a specific referral rate or number needed to treat (NNT), and the model is recalibrated annually. Outcomes such as change in total cholesterol, referral completion, or cardiovascular events are measured six months later.

\begin{table*}[ht]
\centering
\footnotesize
\begin{tabularx}{\textwidth}{lXX}
\toprule
\textbf{Design Element} & \textbf{General Setting} & \textbf{Cardiovascular Risk Prediction Example} \\
\midrule
\textbf{Inclusion criteria} & Patients enrolled in the trial & Adults aged 40--79 \\
\textbf{Exclusion criteria} & Patients not eligible for trial & History of cardiovascular disease or contraindication to preventive therapy \\
\textbf{Model} & Predictive model for adverse events & Model predicting 10-year atherosclerotic cardiovascular risk \cite{goff20142013} \\
\textbf{Intervention} & Action triggered for patients above risk threshold & Referral to cardiovascular prevention program \\
\textbf{Comparator} & Standard care for patients below threshold & Routine care without referral \\
\textbf{Trigger event} & Event that initiates model use & Annual primary care visit \\
\textbf{Allocation} & How high-risk patients are assigned to intervention & Automatic referral if risk exceeds a threshold \\
\textbf{Threshold adaptation} & Criteria for setting threshold & Threshold adjusted to target referral rate or number needed to treat \\
\textbf{Model adaptation} & Strategy for model updates & Annual recalibration or revision using recent data \\
\textbf{Follow-up period} & Observation window after trigger & Six months after the index visit \\
\textbf{Outcome} & Measured effect of intervention & Change in cholesterol, referral completion, or cardiovascular event \\
\bottomrule
\end{tabularx}
\caption{\textit{Key elements of an adaptive regression discontinuity (RD) design.} 
A risk prediction model guides intervention decisions, with both the model and threshold allowed to evolve over time. We illustrate key elements in general terms and through a cardiovascular risk example.}
\label{tab:rd_emulation}
\end{table*}

As the trial unfolds, each patient contributes data for evaluating the intervention. The $i$th patient to reach the trigger event presents with covariates $X_i$, which feed into the current model to produce a predicted risk score $R_i$. This is compared to a threshold $C_i$ to determine whether to initiate the intervention ($A_i = 1$) or comparator ($A_i=0$). If $R_i \geq C_i$, $A_i = 1$; otherwise, $A_i = 0$. The outcome $Y_i$ is observed later. For notational simplicity, we shift the risks so that $C_i = 0$.

We embed these variables in the causal framework of Richardson and Robins \cite{richardson2013single}, which unifies two major traditions in causal inference: potential outcomes and Pearl's do-calculus. The framework specifies how each variable is generated (see \textit{Materials and Methods}).  For example, a patient's predicted risk depends on their covariates and possibly on past patients' data, their intervention assignment depends on predicted risk, and their outcome depends on both the patient's treatment assignment and their covariates. 

This framework defines \textit{potential outcomes}: hypothetical versions of each variable if we had intervened to assign specific treatments to patients up to that point. We generate these by modifying the original model to replace the observed treatment decisions with fixed values. A key property, causal irrelevance, then lets us equate these potential outcomes to other versions of the potential outcomes and to the original variables we observe (Lemma \ref{propcausalirr}) \cite{malinsky2019potential}.

Relationships among the original variables are visualized with a directed acyclic graph (DAG), where each node represents a variable and arrows indicate how variables influence one another. Relationships among potential outcomes are visualized with a single-world intervention graph (SWIG). Figure~\ref{fig:dagscm} shows both the DAG and SWIG for a study with three patients. Since the model or threshold may adapt over time, we allow a patient's predicted risk to depend on data from earlier patients. This introduces \textit{interference}, as treatment decisions for one patient may affect the potential outcomes of those who follow. In the SWIG, this appears in terms like $R_2(a_1)$ and $R_3(a_{1:2})$, which represent predicted risks under hypothetical intervention assignments to earlier patients. While there is interference in risk predictions, causal irrelevance guarantees that a patient's outcome under hypothetical treatment assignments depends only on their own treatment assignment---e.g., $Y_3(a_{1:3})$ simplifies to $Y_3(a_3)$.

\begin{figure}[!ht]
\centering
\scalefont{1}
\tikzset{elliptic state/.style={draw,ellipse}}
\begin{tikzpicture}[scale=1]
\begin{scope}
    \node at (-2,1) {\Large \textbf{A}};
    \node[state] (x1) at (0.2,0) {$X_1$};
    \node[state] (r1) at (2.2,0) {$R_1$};
    \node[state] (a1) at (4.2,0) {$A_1$};
    \node[state] (y1) at (6.2,0) {$Y_1$};
    \path (x1) edge (r1);
    \path (r1) edge (a1);
    \path (a1) edge (y1);
    \node[state] (x2) at (0.2,-2) {$X_2$};
    \node[state] (r2) at (2.2,-2) {$R_2$};
    \node[state] (a2) at (4.2,-2) {$A_2$};
    \node[state] (y2) at (6.2,-2) {$Y_2$};
    \path (x2) edge (r2);
    \path (r2) edge (a2);
    \path (a2) edge (y2);
    \node[state] (x3) at (0.2,-4) {$X_3$};
    \node[state] (r3) at (2.2,-4) {$R_3$};
    \node[state] (a3) at (4.2,-4) {$A_3$};
    \node[state] (y3) at (6.2,-4) {$Y_3$};
    \path (x3) edge (r3);
    \path (r3) edge (a3);
    \path (a3) edge (y3);
    \path (x1) edge[black] (r2);
    \path (y1) edge[black] (r2);
    \path (x2) edge[black] (r3);
    \path (y2) edge[black] (r3);
    \path (x1) edge[out=25,in=155,looseness=1.2] (y1);
    \path (x2) edge[out=25,in=155,looseness=1.2] (y2);
    \path (x3) edge[out=25,in=155,looseness=1.2] (y3);
    \path (x1) edge[black,out=205,in=170] (r3);
   \path (y1) edge[black,out=330,in=15, looseness=1.47] (r3);
    \path (x1) edge[black] (r2);
\end{scope}
\begin{scope}[yshift=-190, xshift=-50]
    \node (B) at (-0.5,1) {\Large\textbf{B}};
    \node[elliptic state] (x1) at (0,0) {$X_1$};
    \node[elliptic state] (r1) at (2.1,0) {$R_1$};
    \node[name=a1,shape=swig vsplit,
        swig vsplit={gap=3pt}] at (5.3,0){
    \nodepart{left}{$A_1$}
    \nodepart{right}{$a_1$} };
    \node[elliptic state] (y1) at (8.7,0) {$Y_1(a_1)$};
    \path (x1) edge (r1);
    \path (r1) edge (a1);
    \path (a1) edge (y1);
    \node[elliptic state] (x2) at (0,-2) {$X_2$};
    \node[elliptic state] (r2) at (2.1,-2) {$R_2(a_1)$};
    \node[name=a2,shape=swig vsplit,
        swig vsplit={gap=3pt}] at (5.5,-2){
    \nodepart{left}{$A_2(a_1)$}
    \nodepart{right}{$a_2$} };
    \node[elliptic state] (y2) at (8.7,-2) {$Y_2(a_2)$};
    \path (x2) edge (r2);
    \path (r2) edge (a2);
    \path (a2) edge (y2);
    \node[elliptic state] (x3) at (0,-4) {$X_3$};
    \node[elliptic state] (r3) at (2.1,-4) {$R_3(a_{1:2})$};
    \node[name=a3,shape=swig vsplit,
        swig vsplit={gap=3pt}] at (5.5,-4){
    \nodepart{left}{$A_3(a_{1:2})$}
    \nodepart{right}{$a_3$} };
    \node[elliptic state] (y3) at (8.7,-4) {$Y_3(a_3)$};
    \path (x3) edge (r3);
    \path (r3) edge (a3);
    \path (a3) edge (y3);
    \path (x1) edge[black] (r2.130);
    \path (y1.220) edge[black] (r2.40);
    \path (x2) edge[black] (r3.130);
    \path (y2.250) edge[black] (r3.40);
    \path (x1) edge[out=22,in=158,looseness=1.1] (y1);
    \path (x2) edge[out=25,in=155,looseness=1.2] (y2);
    \path (x3) edge[out=25,in=155,looseness=1.2] (y3);
    \path (x1) edge[black,out=225,in=170] (r3);
   \path (y1) edge[black,out=320,in=10, looseness=1.4] (r3);
\end{scope}
\end{tikzpicture}
\caption{\textit{Graphs depicting interference from adapting risk prediction models and thresholds over time.} \textbf{A)} Directed acyclic graph (DAG) induced by assumed structural causal model when only three patients are considered. \textbf{B)} Corresponding single world intervention graph (SWIG). Earlier patients can shape the predicted risk of later patients, e.g. $Y_1$ might affect $R_3$. This creates interference, seen in the SWIG as potential outcomes that depend on past interventions, like $R_3(a_{1:2})$. }
\label{fig:dagscm}
\end{figure}

\subsection*{Causal effect identification and estimation} To assess the causal influence of an intervention for patient $i$, our causal estimand of interest is
\begin{align*}
\beta_i(r_i) = \E[Y_i(1) - Y_i(0) \mid R_i = r_i], 
\end{align*}
capturing the local average treatment effect (ATE) of the intervention for patient $i$ given a specific risk score $r_i$. A key distinction from typical RD designs is that risk scores $R_i$ are not assigned independently. Instead, they depend on a prediction model and threshold that update as each new patient is evaluated. Consequently, $R_i$ is influenced not just by patient $i$'s characteristics but also by previously observed data and decisions. This interference results in a \textit{patient-specific} estimand, making it unclear how to leverage data from other patients for estimation. 

One idea is to condition directly on covariates:
\begin{align*}
\E[Y_i(1) - Y_i(0) \mid X_i].
\end{align*}
Since we assume $(X_i, Y_i(0), Y_i(1))$ are i.i.d. across individuals, this estimand can be estimated using data from others with similar covariates. Averaging over $X_i$ given $R_i = r_i$ then yields $\beta_i(r_i)$. In practice, however, the more covariates we condition on, the harder it becomes to find patients with similar covariates, degrading estimation quality.

We propose a strategy that permits borrowing information from other patients without overly restricting the data we can use. Rather than relying only on patient $i$'s actual risk score, we examine what their risk would have been under the models and thresholds used for earlier patients. We term this concept  \textit{counterfactual risk} and denote it by $\bar{R}_{i,j}$ ($j \leq i$), representing the risk score patient $i$ would have obtained using the model and threshold from patient $j \leq i$. When $j = i$, the counterfactual risk $\bar{R}_{i,i}$ is simply $R_i$.

Our causal estimand can be expressed in terms of these counterfactual risks:
 \begin{align*}
    \beta_i(r_i) := \E\left[ \E\left[ Y_{i}(1)-Y_{i}(0) \mid \bar{R}_{i,1:i} \right] \mid R_i = r_i \right],
\end{align*} 
where the inner expectation now measures the average effect of the intervention for patient $i$ given their counterfactual risk scores. In the special case where the prediction model and threshold do not change over time, each patient is evaluated using the same model and threshold. As a result, all counterfactual risks collapse to the observed risk score $R_i$, and the estimand simplifies to the standard RD causal effect.

In this form, this estimand allows us to pool information across patients, contingent upon two key assumptions. First, we require a continuity condition: the conditional mean of $Y_i(1)$ and $Y_i(0)$ must vary continuously in the counterfactual risks $R_{i,1:i}$, extending the classic RD continuity condition to accommodate adapting models and thresholds (Assumption~\ref{assm:continuity}). This ensures that patients with similar counterfactual risks yield similar expected potential outcomes. Second, we assume that we can reconstruct the counterfactual risks that would arise if patients $k$ and $i$ swapped positions in the sequence (Assumption~\ref{assm:exchangeability}). The resulting counterfactual risks for patient $k$ are denoted by $\bar{R}_{k,1:i}$, with $\bar{R}_{k,k} = R_k$.

When we hypothetically swap these patients, our conditioning event now involves $R_k$. Because treatment is assigned deterministically based on this risk, this also fixes their treatment status. As a result, we are conditioning not only on risks but also on treatment. This allows us to invoke the principle of consistency, replacing the potential outcome $Y_k(a)$ with the observed outcome $Y_k$. By conditioning on the counterfactual risks for all patients $k \leq i$, we can use each patient's observed data to inform estimation of the causal effect for patient $i$. These observations motivate the following theorem, proved in \textit{Materials and Methods}:
\begin{theorem}\label{thm:iden}
Fix a patient $k \leq i$, and consider a risk profile $r_{1:i}$ and a direction $d_{1:i}$. Define $a = 1$ if $r_k > 0$ or if $r_k = 0$ and $d_k > 0$; otherwise, set $a = 0$. Assuming continuity (Assumption~\ref{assm:continuity}) and the ability to compute counterfactual risks (Assumption~\ref{assm:exchangeability}), then 
\begin{align*}
\E[Y_i(a) \mid \bar{R}_{i,1:i} = r_{1:i} ] = \lim_{h \to 0^+} \E\left[ Y_k \mid \bar{R}_{k,1:i} = r_{1:i} + h \, d_{1:i} \right]
\end{align*}
provided these conditional expectations are defined.
\end{theorem}
Theorem~\ref{thm:iden} shows that the expected potential outcome for patient $i$ under treatment $a$ can be approximated using the observed outcomes of earlier patients who received treatment $a$ and had similar counterfactual risks. We denote this conditional mean by $m_a(r_{1:i})$. After learning this function, we average $m_a(\bar{R}_{i,1:i})$ over all components of the risk profile except the observed risk score $R_i = r_i$. Because counterfactual risks are constructed to follow the same distribution across patients, we can do this averaging using counterfactual risks from earlier patients $k\leq i$. This yields the expected potential outcome under each treatment condition at that risk level, $\mu_a(r_i)$, from which we compute the causal effect: $\beta(r_i) = \mu_1(r_i)-\mu_0(r_i)$.

Guided by this result, our estimator for $\beta_i(r_i)$ proceeds in three steps. First, we model the observed outcome $Y_k$ as a function of the full counterfactual risk profile $\bar{R}_{k,1:i}$, using a generalized linear model (GLM) with separate spline terms for the two intervention groups. Although $\bar{R}_{k,1:i}$ appears high-dimensional (it has $i$ entries), it typically has low intrinsic dimension. For example, if models evolve slowly over time or if risk thresholds are updated but models are not, much of the variation in $\bar{R}_{k,1:i}$ can be captured by a small number of components. To capture this structure, we residualize the risks on $\bar{R}_{k,i}$ and then reduce dimensionality with principal component analysis (PCA). Second, to approximate the conditional expectation in the identification formula, we use the fitted GLM to predict potential outcomes under each treatment condition. Finally, we apply kernel smoothing over the counterfactual risk vectors, weighting patient $k$ according to how close their $\bar{R}_{k,i}$ is to $r_i$. This marginalization yields an estimate of $\beta_i(r_i)$. Full details are in \textit{Materials and Methods}.

\subsection*{Simulation} \textit{Increasing uptake of a resource-limited program (Scenario 1):} We simulate a program that refers patients to preventive care if their predicted 10-year ASCVD risk exceeds a threshold \cite{goff20142013}. To reflect resource constraints, the program can only handle a 30\% referral rate. So, the risk threshold is adapted over time based on observed risks.

\begin{figure*}[ht!]
\centering
    \includegraphics[width=\textwidth]{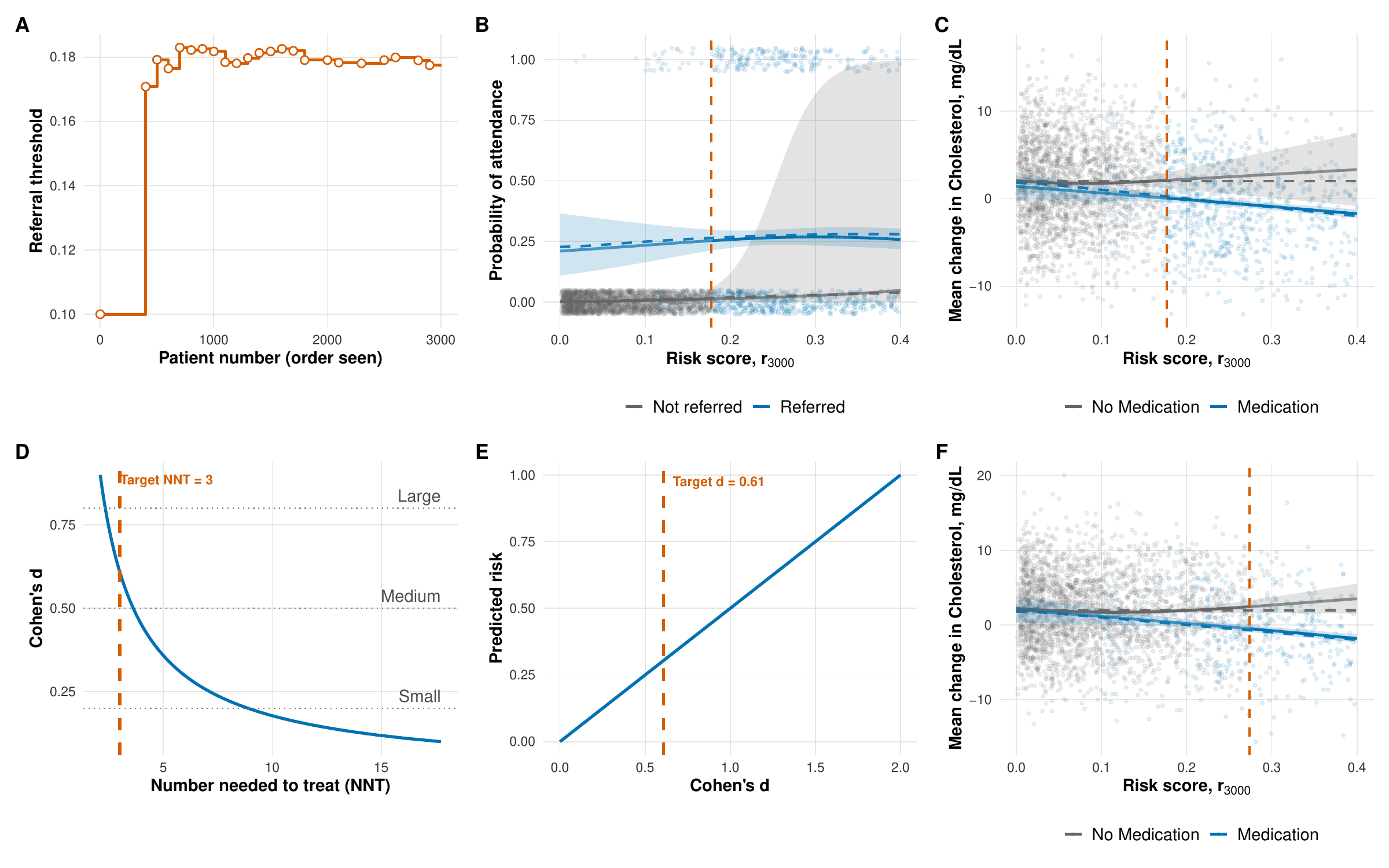}
    \caption{\textit{Visual summary of three simulation scenarios focused on adapting the risk threshold.}
\textbf{A)} Example trajectory of risk threshold over time (Scenario 1).
\textbf{B)} Estimated (solid) and true (dashed) conditional mean potential outcomes for program attendance under referral and non-referral, plotted as functions of the risk score $r_{3000}$ (Scenario 1).
\textbf{C, F)} Estimated (solid) and true (dashed) conditional mean potential outcomes for change in total cholesterol under mediation and no medication, plotted as functions of the risk score $r_{3000}$ (Scenarios 2 and 3, respectively).
\textbf{D)} Mapping from  number needed to treat to Cohen's $d$ (Scenario 3).
\textbf{E)} Mapping from Cohen's $d$ to predicted risk (Scenario 3).
In \textbf{B}, \textbf{C}, and \textbf{F}, shaded regions indicate 95\% confidence intervals, vertical lines denote the last risk threshold, and each dot represents a patient.}
    \label{fig:threshold_adaptation}
\end{figure*}

We generate realistic patient covariates by sampling from the National Health and Nutrition Examination Survey (NHANES) 2017–March 2020 pre-pandemic data files \cite{stierman2021national}. The first 400 patients are referred if their risk exceeds a threshold of .10; thereafter, the threshold is updated every 100 patients to target a 30\% referral rate. Figure~\ref{fig:threshold_adaptation}A shows the threshold rising and stabilizing around .18,  reflecting the shift needed to meet this target in the simulated population.

Each patient's outcome is whether they attend a preventative care visit. This probability depends on both their baseline risk and whether they are referred, with attendance most likely among referred patients at moderate risk  (around 25\%).  Our estimator recovers the conditional mean potential outcomes, $\mu_1(r_i)$ and $\mu_0(r_i)$ for $i=3000$, representing the expected probability of attendance under referral and non-referral, respectively, as functions of the risk score $r_i$. Figure~\ref{fig:threshold_adaptation}B compares these estimated curves to the true conditional means used to generate the data. Our estimated curves tracks the true curves closely, with greater uncertainty in regions with limited data (e.g., low-risk referred patients). Evaluated at the final risk threshold, we estimate a local ATE of .240 (95\% CI: [.193, .288]); the true effect is .247.

\textit{Prioritizing patients for a new lipid-lowering medication (Scenario 2):} We next simulate a pragmatic trial of a new lipid-lowering medication, again using an adaptive risk threshold to guide treatment. The outcome is now continuous: change in total cholesterol from baseline to follow-up. In our simulation, untreated patients experience modest increases in cholesterol over time, while treated patients show decreases, with larger reductions among those at higher risk. Figure~\ref{fig:threshold_adaptation}C shows our estimated conditional means, $\mu_1(r_i)$ and $\mu_0(r_i)$ for $i=3000$, which represent the expected change in cholesterol under medication and no medication, respectively. At the final threshold, we estimate a local ATE of –2.035 mg/dL (95\% CI: [–3.063, –1.006]); the true effect is –1.742 mg/dL.

\textit{Balancing benefit and burden through number needed to treat (Scenario 3):} In this scenario, we simulate an adaptive threshold that offers lipid-lowering medication to help lower cholesterol, but only when the expected benefit exceeds a target number needed to treat (NNT). Because NNT is typically defined for binary outcomes, we translate it of $3$ into an equivalent effect size (Cohen's $d$) suitable for continuous outcomes, using an established conversion formula \cite{furukawa2011obtain}. Our target is a NNT of 3, which implies 2 out of 3 treated patients are expected to benefit from medication and translates into a Cohen's $d$ of $0.61$ (Figure~\ref{fig:threshold_adaptation}D). Figure~\ref{fig:threshold_adaptation}E plots predicted risk against the true effect size, based on the simulated outcome model. The target $d = 0.61$ maps to a predicted risk of .305, which serves as our ideal treatment threshold.

The first 400 patients are treated using a fixed threshold of .10. Thereafter, the threshold is updated every 100 patients using our estimator, which learns effects across risk levels, converts them to Cohen's $d$, and selects the threshold that best matches the target. To stabilize decisions, each new threshold is smoothed with the prior one, yielding a final threshold of .274 in our simulation, slightly below the ideal .305. Figure~\ref{fig:threshold_adaptation}C plots estimated conditional mean curves, $\mu_1(r_i)$ and $\mu_0(r_i)$ for $i=3000$, representing the expected change in cholesterol under medication and no medication. At the last threshold, we estimate a local ATE of –2.954 mg/dL (95\% CI: [–3.902, –2.007]); the true effect is –2.696. 

\begin{figure}[ht!]
\centering
    \includegraphics[width=.56\textwidth]{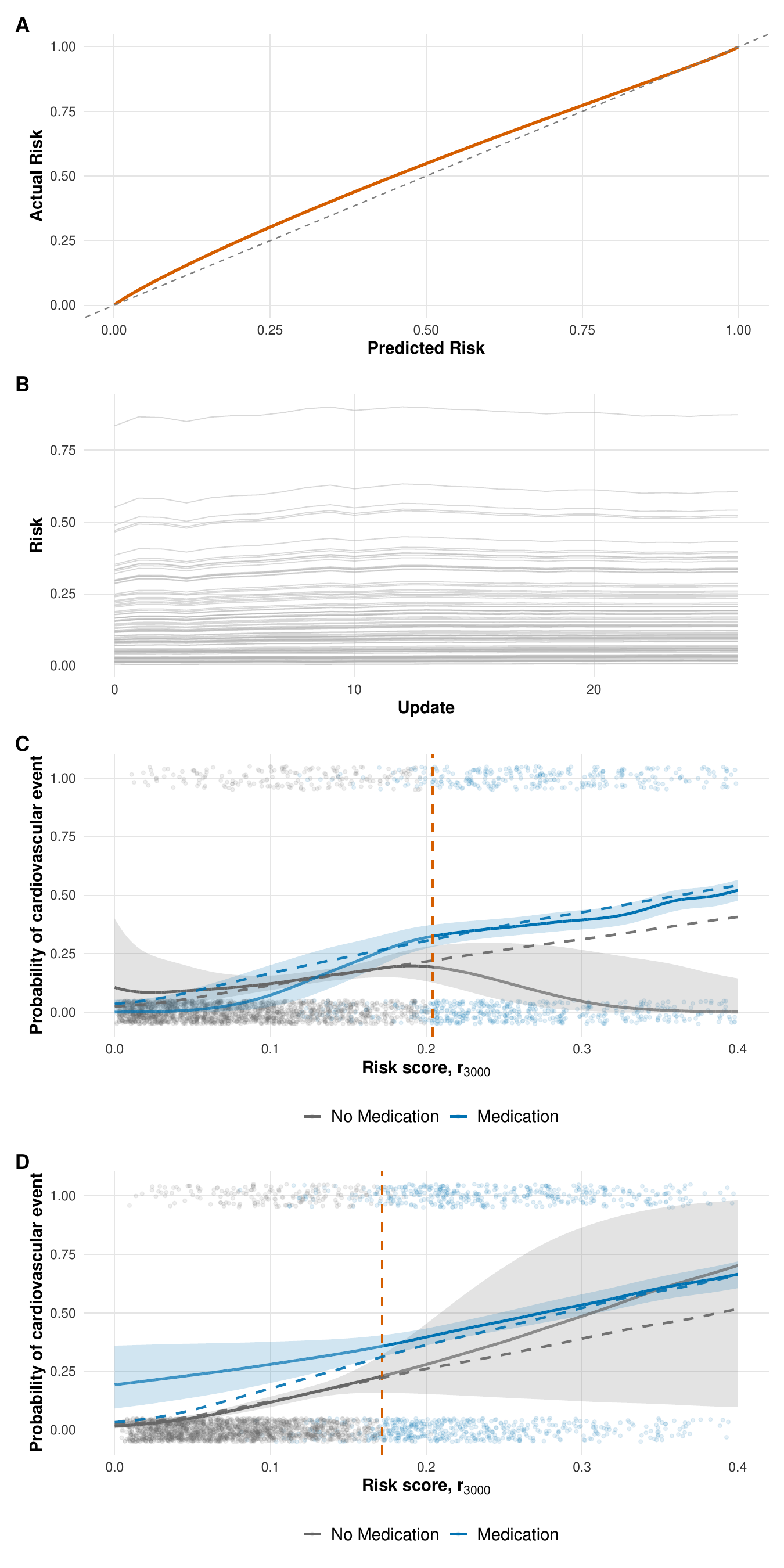}
    \caption{\textit{Visual summary of two simulation scenarios involving adaptation of the risk prediction model.}
\textbf{A)} Simulated mismatch between predicted and actual risk (Scenario 4).
\textbf{B)} Trajectory of predicted risk for the first 100 patients across model updates as it recalibrates using prior patient data (Scenario 4).
\textbf{C, D)} Estimated (solid) and true (dashed) conditional mean potential outcomes for ASCVD event under mediation and no medication as function of the risk score $r_{3000}$ (Scenario 4 and 5, respectively). 
In \textbf{C} and \textbf{D}, shaded regions are 95\% confidence intervals, vertical dashed lines show the final risk threshold, and each dot represents a patient.}
    \label{fig:model_adaptation}
\end{figure}

\textit{Recalibrating the risk model (Scenario 4):} In this and the next scenario, we shift focus from adapting thresholds to adapting the prediction model itself. The outcome is ASCVD events, and the model is recalibrated to better align predicted and actual risks. Figure~\ref{fig:model_adaptation}A shows the initial miscalibration. After a warm-up period, the model is recalibrated every 100 patients, with updates smoothed over time. Figure~\ref{fig:model_adaptation}B illustrates how recalibrated risk scores evolve for the first 100 patients. While the estimated curves for $\mu_1(r_i)$ and $\mu_0(r_i)$ with $i=3000$, recover the overall risk–response pattern, they exhibit greater error than in previous scenarios (Figure~\ref{fig:model_adaptation}C). At the final threshold, the estimated local ATE is .132 (95\% CI: [.039, .225]), compared to a true effect of .090.

\textit{Revising model to remove race and gender (Scenario 5)} The final scenario reflects a growing trend to revise clinical prediction models by removing demographic variables. We update the model coefficients to exclude race and gender. As shown in Figure~\ref{fig:model_adaptation}D, our estimator still captures the key relationship between risk and mean potential outcomes. At the final threshold, the estimated local ATE is .128 (95\% CI: [.034, .221]), versus a true effect of .090.

\begin{figure*}[ht!]
\centering
    \includegraphics[width=\textwidth]{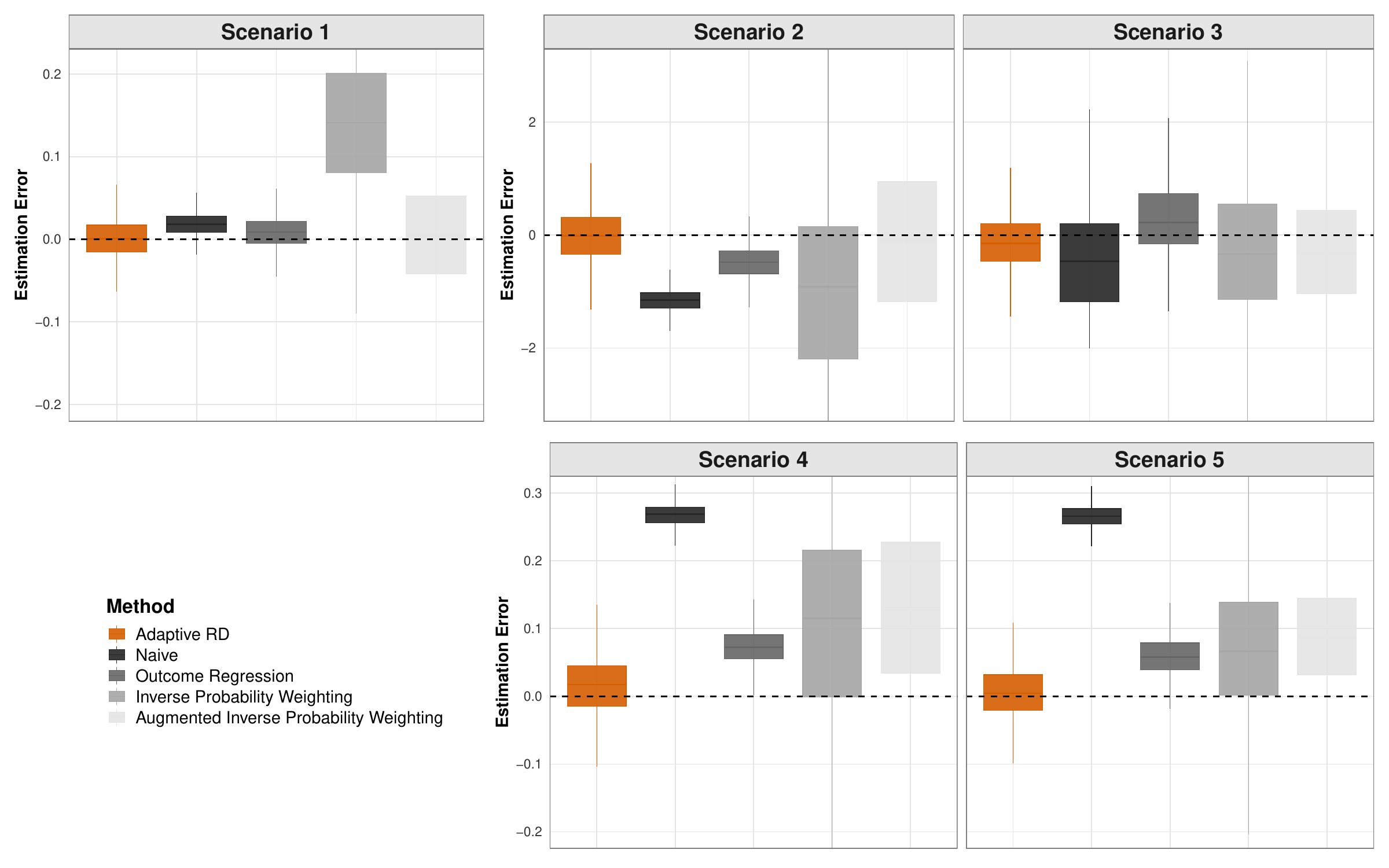}
    \caption{\textit{Comparison of estimation error in the local ATE at the final risk threshold across five simulation scenarios.} Each boxplot shows the distribution of estimation error over 2000 replications. Our adaptive RD estimator is compared against a naive difference in means, outcome regression, inverse probability weighting (IPW), and augmented IPW.}
    \label{fig:comparison}
\end{figure*}

\textit{Performance relative to comparators:}
We repeated each simulation 2000 times and compared our estimator to four alternatives. Figure~\ref{fig:comparison} shows that our adaptive RD estimator consistently yields lower bias and mean squared error.

\section*{Discussion}

We developed a quasi-experimental framework based on RD designs to evaluate how risk prediction models perform when paired with specific interventions in real-world care. These models increasingly guide clinical decisions. Yet, the very settings in which they are deployed (where models are retrained, revised, and adapted) can make formal evaluation difficult. Our framework fills a critical gap: the ability to rigorously evaluate model-intervention pairings as they adapt in response to accumulating data.

Allowing models and thresholds to evolve introduced statistical challenges that standard RD designs do not address. Chief among these is \textit{interference}: as outcomes from past patients are observed, they influence future treatment decisions via updates to the model or threshold. This violates the stable unit treatment value assumption \cite{rubin1980randomization}, breaks the usual i.i.d. assumption, and complicates identification. While some strategies allow for interference in RD designs \cite{cattaneo2016inference,rosenbaum2007interference,aronow2017regression}, they do not estimate the local ATE. Rather, they estimate either only confidence intervals or effects that depend on how the intervention is assigned \cite{hudgens2008toward}. Further, they do not take advantage of the specific form of interference created by adapting models and thresholds, and so may be too general to be effective in this context. Interference remains an issue even if only the cutoff is adapted, since methods for managing multiple cutoffs do not permit interference \cite{cattaneo2016interpreting,butler2009regression,behaghel2008perverse}. 

The central innovation of our approach is the use of \textit{counterfactual} risks: the risks each patient would have been assigned under hypothetical reorderings of the patients. By conditioning on these risks, we restore local exchangeability, recover the treatment actually received, and invoke consistency to connect observed outcomes to the potential outcomes. Our estimator builds on this by leveraging the fact that model and threshold updates, though unconstrained, are typically sparse or gradual. This implies that the vector of counterfactual risks lies approximately in a low-dimensional space, even when derived from high-dimensional covariates. For example, if the model is updated once, counterfactual risks span a two-dimensional space: those from before and after the update. We first reduce the problem to this lower-dimensional representation, and then model the outcome as a smooth function of treatment and counterfactual risks.

Though illustrated through specific examples, the framework is designed to be plug-and-play: each design component in Table~\ref{tab:rd_emulation} can be modified to fit alternative settings and operational goals. For example, thresholds may be adapted using different targeting rules, models updated through other data-driven strategies \cite{steyerberg2004validation,janssen2008updating,binuya2022methodological,su2018review}, and outcomes generalized to any GLM-compatible form. This flexibility supports broad application across use cases. In fact, our original motivation came from a different setting: a study of fall prevention in older adults, where evolving risk models embedded in the ED workflow guide referrals to a prevention clinic \cite{patterson2017using,patterson2018using,patterson2019training,jacobsohn2022collaborative,hekman2023effect,hekman2024dashboarding}. 

Our experience simulating the adaptive RD design surfaced several useful lessons. Adaptation can improve estimation by introducing variation in treatment assignment near the threshold, which creates more informative contrasts between treated and untreated patients. For example, a patient with a 12\% predicted risk might receive treatment early in the study when the threshold is 10\%, but go untreated later when it rises to 15\%. Adaptation can also align treatment decisions with operational realities, such as limiting referrals when specialist access is constrained or removing race-based adjustments, as in recent changes to the estimated glomerular filtration rate equation \cite{delgado2021unifying}. That said, adaptation brings risks. Adapting too early or too often can cause models or thresholds to chase random noise rather than true patterns. This not only risks misleading estimates but can also undermine identification, as a single patient might overly influence future updates, complicating estimation of their counterfactual risks. We recommend delaying adaptation with a warm-up period, shrinking updates toward prior values, and limiting how often updates occur. Finally, simulation is a valuable tool for stress-testing proposed designs under realistic conditions. Our open-source codebase supports tailoring these simulations to the needs of individual healthcare systems.

Several limitations warrant discussion. First, our framework estimates the effect of treatment assignment based on predicted risk, not actual treatment received. With imperfect compliance, the estimand reflects an intention-to-treat effect. Second, like all RD designs, we assume individuals cannot manipulate treatment assignment. This means patients or clinicians cannot systematically alter model inputs to influence treatment. Violations could bias estimates. Third, we illustrate possible adaptation strategies but do not provide full guidance on how to design or tune them. We only caution that adapting too aggressively can distort estimates and threaten identification. Fourth, we used simple imputation for missing data. More robust approaches should be embedded in the prediction model itself, where imputation and risk estimation are jointly handled. Fifth, we focused on continuity-based identification, but the framework could be extended to use a local randomization assumption, which may be more natural in some settings \cite{lee2008randomized,cattaneo2016inference}. Finally, RD designs generally need more patients than RCTs, since only those near the threshold contribute to each estimate. Still, our simulations showed that 3,000 patients yielded stable results, and this scale is often feasible in health systems.

\section*{Materials and Methods}

\subsection*{Causal framework} We associate each patient $i = 1, \dots, n$ with several random variables:
\begin{itemize}
\item Covariates $X_i$ (model input)
\item Predicted risk $R_i$ (model output)
\item Risk threshold $C_i$
\item Intervention assignment $A_i$ ($1$ if assigned, $0$ otherwise)
\item Outcome $Y_i$
\end{itemize}
Patients are indexed by their order of reaching the trigger event. Predicted risk $R_i$ is computed from covariates $X_i$ using the model available at that time. Treatment assignment follows $A_i = \mathbb{I}(R_i \geq C_i)$. The model and threshold can evolve without constraints on timing or update approach. Without loss of generality, and for our theoretical exposition, we redefine the risk score $R_i$ so that thresholds are always zero ($C_i = 0$).

We embed these variables within the causal framework of Richardson and Robins \cite{richardson2013single}, which blends the versatility of potential outcomes with the graphical and axiomatic tools of Pearl's do-calculus. We assume a \textit{structural causal model} to describe the sequential generation of variables:
\begin{align*}
   X_i &:= \epsilon_i  \\
  R_i &:= f_{r_i}(X_i,X_{1:i-1},Y_{1:i-1}) \\
  A_i &:= \mathbb{I}(R_{i}\geq 0) \\
  Y_i &:= f_{y}(X_i,A_i,\eta_i)
\end{align*}
Here, $f_{r_i}$ and $f_y$ are functions assigning predicted risks and outcomes. Terms $\epsilon_i$ and $\eta_i$ represent independent random noise, iid across patients. These assumptions imply covariates $X_i$ are iid, while $R_i$, $A_i$, and $Y_i$ are generally not. This dependence arises because predicted risks $R_i$ depend on past patients' covariates and outcomes, capturing model adaptations. (Technically, $R_i$ could also depend on past treatment assignments and past risk scores, but since these are deterministic functions of $X_{1:j}$ and $Y_{1:j-1}$, we omit them from $f_{r_i}$ for notational simplicity.) However, outcomes $Y_i$ depend solely on the patient's covariates, treatment, and independent noise, ensuring $Y_i$ are conditionally iid given $A_i$ and $X_i$. The implied DAG when $n=3$ is summarized in Figure~\ref{fig:dagscm}A.

Using this structural causal model, we define potential outcomes for subsets $B \subseteq \{1,\dots,n\}$ and intervention vectors $a_{1:n} \in \{0,1\}^n$ sequentially from $i = 1$ to $n$:
\begin{align*}
  X_i(a_B) &:= \epsilon_i  \\
  R_i(a_B) &:= f_{r_i}(X_i(a_B), X_{1:i-1}(a_B),Y_{1:i-1}(a_B)) \\
  A_i(a_B) &:= \mathbb{I}(R_{i}(a_B)\geq 0) \\
  Y_i(a_B) &:= f_{y}(X_i(a_B),\bar{A}_i(a_B),\eta_i),
\end{align*}
where 
\begin{align*}
  \bar{A}_i(a_B) &:= \begin{cases}
    a_i & i \in B \\
    A_i(a_B) & i \notin B
  \end{cases}.
\end{align*}
This approach, from Richardson and Robins \cite{richardson2013single}, defines potential outcomes by substituting intervention assignments $A_i$ with fixed values $a_i$. The resulting variables and intervened values form a single-world intervention graph (SWIG). It can derived by splitting each intervened node in the original DAG into two: one part retains incoming edges and remains stochastic; the other takes the fixed intervention value and inherits the outgoing edges. 

This construction directly implies consistency and causal irrelevance, eliminating the need for separate assumptions. Causal irrelevance means assigning an intervention $a_i$ does not affect any variable in the SWIG that is not a descendant of $a_i$, simplifying potential outcomes by removing unnecessary arguments \cite{malinsky2019potential}. For example, $Y_i(a_{1:n})$ is not a descendant of any $a_j$ for $j \ne i$ in the SWIG, leading to the simplification $Y_i(a_{1:n}) = Y_i(a_i)$. These simplifications are formalized:
\begin{lemma}\label{propcausalirr}
For all $i$:
$X_i(a_{1:n}) = X_i$,
$R_i(a_{1:n}) = R_i(a_{1:i-1})$,
$A_i(a_{1:n}) = A_i(a_{1:i-1})$, and
$Y_i(a_{1:n}) = Y_i(a_i)$.
\end{lemma}
This lemma explicitly characterizes interference in our design. Intervention assignments do not affect outcomes for later patients, while they do influence risk predictions and treatment decisions, but only through prior assignments. When we apply this lemma with $n=3$ patients, we obtain the SWIG in Figure~\ref{fig:dagscm}B.

Because $Y_i(a_{1:n}) = Y_i(a_i)$, we use the shorthand $Y_i(a)$ to denote the outcome for patient $i$ under intervention $a$, assuming all other treatments remain unchanged. While this notation does not explicitly indicate that only patient $i$ is intervened upon, we adopt it throughout and rely on the subscript in the outcome to implicitly specify the target of the intervention.

\subsection*{Target causal effect}  We consider causal effects at the level of individual patients as they arrive sequentially to the system. Because the risk prediction model and decision threshold may evolve over time, the relationship between risk and treatment benefit must be defined relative to each patient's position in the sequence. Accordingly, we define a patient-specific local average treatment effect (ATE) for patient $i$ when their predicted risk is fixed at a specific value $r_i$:
 \begin{align*}
    \beta_i(r_{i}) := \E\left[ Y_i(1) - Y_i(0) \mid \bar{R}_{i} = r_{i} \right]
\end{align*} 
whenever this conditional expectation is defined.

We define the counterfactual risk of patient $i$ under the model–threshold pair applied to patient $j$ as
\begin{align*}
\bar{R}_{i,j} := f_{r_j}(X_i, X_{1:j-1}, Y_{1:j-1})
\end{align*}
for $j \leq i$. This quantity represents the risk score that patient $i$ would have received if they had been the $j$th person in the sequence. Notice the observed risk for patient $i$ is simply $\bar{R}_{i,i}=R_i$.

To identify $\beta_i(r_{i})$, we introduce two key assumptions. First, we require a continuity assumption analogous to that of standard RD designs \cite{hahn2001identification}, but generalized for scenarios where the prediction model and thresholds adapt over time. Formally:
\begin{assm} \label{assm:continuity}
For each $a \in \{0,1\}$, $\E[Y_i(a) \mid \bar{R}_{i,1:i} = r_{1:i}]$ is continuous in the vector $r_{1:i}$. 
\end{assm}

Our second assumption ensures that counterfactual risks can be computed under hypothetical reordering of patients. For indices $k \geq j$, swapping patients $i$ and $k$ transforms $\bar{R}_{i,j}$ to
\begin{align*}
f_{r_j}(X_k,X_{1:j-1},Y_{1:j-1}) := \bar{R}_{k,j},
\end{align*}
since $X_{1:j-1}$ and $Y_{1:j-1}$ remain unchanged, while $X_i$ becomes $X_k$. This represents the risk score patient $k$ would have received under model and threshold $j$, and is fully determined by observed variables.

For $k < j$, the definition is more subtle. In our structural causal model, each variable is a deterministic function of the underlying noise terms $(\epsilon_i, \eta_i)$. Swapping patients $i$ and $k$ corresponds to exchanging their respective noise terms. Let $g$ denote the function such that $\bar{R}_{i,j}=g(\epsilon_{1:i}, \eta_{1:i})$. Then, applying a permutation $\pi$ that swaps indices $i$ and $k$ yields:
\begin{align*}
 g(\epsilon_{\pi(1:i)}, \eta_{\pi(1:i)}) := \bar{R}_{k,j}.
\end{align*}
Unlike the case when $k \geq j$, this expression may involve unobserved terms, depending on how the model and threshold update over time. To ensure identifiability, we assume:
\begin{assm} \label{assm:exchangeability}
For each $j < k < i$, the counterfactual risk $\bar{R}_{k,j}$ can be determined from observed variables $(X_{1:i}, Y_{1:i})$.
\end{assm}

\subsection*{Identification Proof} To prove Theorem~\ref{thm:iden}, fix a patient $k \leq i$, and consider a risk profile $r_{1:i}$ and a direction $d_{1:i}$. Define $a=1$ if $r_k > 0$ or if $r_k = 0$ and $d_k > 0$; otherwise, set $a = 0$. The motivation behind this choices will become clear shortly. Assume all relevant conditional expectations are defined. 

By our continuity assumption (Assumption~\ref{assm:continuity}), we can express the conditional mean of the potential outcome as a limit:
\begin{align*}
\lim_{h \to 0^+} \E[Y_i(a) \mid \bar{R}_{i,1:i} = r_{1:i} +  h \, d_{1:i}].
\end{align*}

Next, we leverage the exchangeability of the noise terms $(\epsilon_i,\eta_i)$ and $(\epsilon_k,\eta_k)$, which are iid by assumption. Every variable appearing in the conditional mean above is a deterministic function of the noise terms. Swapping $(\epsilon_i,\eta_i)$ and $(\epsilon_k,\eta_k)$ --- while holding all other noise terms fixed --- preserves the joint distribution of the noise terms, yielding:
\begin{align*}
&\E[Y_i(a) \mid \bar{R}_{i,1:i} = r_{1:i} +  h \, d_{1:i}] = \\
&\E[Y_k(a) \mid \bar{R}_{k,1:i} = r_{1:i} +  h \, d_{1:i}].
\end{align*}
In particular, swapping the noise terms transforms $Y_i(a) = f_y(X_i,a,\eta_i)$ into $f_y(X_k,a,\eta_k) = Y_k(a)$, since $X_i=\epsilon_i$ and $X_k = \epsilon_k$. Similarly, $\bar{R}_{i,j}$ transforms into $\bar{R}_{k,j}$ by construction: our definition of counterfactual risks ensures this mapping under such a swap. By assumption, $\bar{R}_{k,j}$ is recoverable from observed variables (Assumption~\ref{assm:exchangeability}), with $\bar{R}_{k,k} = R_k$.

Our choice of $a$ and $d_k$ guarantees that, for sufficiently small $h > 0$, the treatment assignment satisfies $A_k = a$ when $R_k = r_k + h d_k$. We can then invoke consistency ($Y_k(a) = Y_k$ whenever $A_k = a$) to obtain:
\begin{align*}
\E[ Y_i(a) \mid \bar{R}_{i,1:j} = r_{1:i} + h \, d_{1:i} ] = \E[ Y_k \mid \bar{R}_{k,1:j} = r_{1:i} + h \, d_{1:i}].
\end{align*}
Substituting this into our earlier expression, we get
\begin{align*}
\E[Y_i(a) \mid \bar{R}_{i:i} = r_{1:i}] = \lim_{h \to 0^+} \E[Y_k \mid \bar{R}_{k,1:i} = r_{1:i} +  h \, d_{1:i}],
\end{align*}
which completes the proof.

\subsection*{Estimation} To estimate $\beta_i(r_i)$, we use data from prior patients $k = 1, \ldots, i$. For each such patient $k$, we approximate counterfactual risk scores $\bar{R}_{k,j}$ by applying $k$'s covariates to the model used for patient $j$, for all $j \le i$. In an abuse of notation, we denote these approximations by $\bar{R}_{k,j}$, even though  they may differ slightly from the true counterfactual risks if patient $k$ contributed to training model $j$. Still, we assume this dependence is negligible, as any individual patient has minimal influence on the model fit.

We then fit a GLM using data from all patients $k = 1, \ldots, i$, regressing the observed outcomes $Y_k$ on the counterfactual risks $\bar{R}_{k,1:i}$. With a specified distributional family (e.g., Bernoulli) and a link function $g(\cdot)$, we model:
$$
g(\mathbb{E}[Y_k \mid \bar{R}_{k,1:i}]) = f_0(\bar{R}_{k,1:i})(1 - A_k) + f_1(\bar{R}_{k,1:i}) A_k,
$$
where $f_0(\cdot)$ and $f_1(\cdot)$ are flexible functions.

Because the predictors $\bar{R}_{k,1:i}$ are high-dimensional and often collinear, we first residualize each $\bar{R}_{k,j}$ (for $j < i$) against the focal risk score $\bar{R}_{k,i}$ using linear regression. That is, for each $j < i$, we regress $\bar{R}_{k,j}$ on $\bar{R}_{k,i}$ across patients $k = 1, \ldots, i$, and retain the residuals. This removes linear dependence on $\bar{R}_{k,i}$, which plays a key role in the target estimand $\beta_i(r_i)$, while preserving variation orthogonal to it. We then apply PCA to the residualized values and use the top components as covariates. Both $f_0(\bar{R}_{k,1:i})$ and $f_1(\bar{R}_{k,1:i})$ are then modeled as natural cubic splines in $\bar{R}_{k,i}$ and a shared linear function of the top principal components of residuals. Knots are chosen separately for each treatment group. 

We estimate the target causal effect $\beta_i(r_i)$ using a kernel-weighted average of predicted treatment effects across prior patients $k = 1, \dots, i$, based on the fitted GLM. Specifically, we compute:
\begin{align*}
\hat\beta_i(r_i) := \sum_{k=1}^i w_k(r_i) \left\{ g^{-1}( \hat{f}_1(\bar{R}_{k,1:i})) -   g^{-1}(\hat{f}_0(\bar{R}_{k,1:i})) \right\},
\end{align*}
where $\hat{f}_1(\cdot)$ and $\hat{f}_0(\cdot)$ are the estimated outcome functions under treatment and control, respectively, and $g^{-1}$ is the inverse link function from the GLM. The weights $w_k(r_i)$ assign higher importance to patients whose counterfactual risk $\bar{R}_{k,i}$ is close to the target $r_i$:
\begin{align*}
w_k(r_i) \propto \exp\left(-(\bar{R}_{k,i} - r_i)^2/(2h^2)\right),
\end{align*}
for a specified bandwidth $h$. By averaging over patients with $\bar{R}_{k,i} \approx r_i$, this estimator implicitly marginalizes over the distribution of the remaining covariates (i.e., the components of $\bar{R}_{k,1:i}$ other than $\bar{R}_{k,i}$), conditional on the focal risk value. 

We apply the delta method to estimate the standard error of $\hat{\beta}_i(r_i)$. We treat the estimator $\hat{\beta}_i(r_i)$ as a smooth function of the regression coefficients $\theta$, defining
$$g(\theta) := \sum_{k=1}^i w_k \left\{ g^{-1}(f_1(R_{k,1:i})) - g^{-1}(f_0(R_{k,1:i})) \right\},$$
where $\theta$ is implicit in the functions $f_1$ and $f_0$. The standard error is then approximated by
$$\widehat{\mathrm{SE}}(\hat{\beta}_i(r_i)) \approx \left[ \nabla_\theta g(\hat\theta)^\top \hat V_\theta \nabla_\theta g(\hat\theta) \right]^{1/2},$$
where $\hat V_\theta$ is the estimated covariance of $\hat\theta$ and $\nabla_\theta g(\hat\theta)$ is the gradient evaluated at the fitted values. A $100(1 - \alpha)\%$ Wald confidence interval is then constructed.

Estimation decisions include GLM family and link function, the degrees of freedom for the splines, the amount of variance captured in PCA, and the kernel bandwidth $h$ used in marginalization.

\subsection*{Simulation}  All simulation code is available in our public repository (\href{https://github.com/cochran4/adaptive_rd}{\texttt{github.com/cochran4/adaptive\_rd}}), with a fully rendered Quarto-based demonstration accessible at \href{https://cochran4.github.io/adaptive_rd/}{cochran4.github.io/adaptive\_rd}.

\textit{Risk prediction model:} We estimated 10-year risk of ASCVD using the Pooled Cohort Equations developed by the American College of Cardiology and American Heart Association \cite{goff20142013}. Risk is predicted separately for sex–race subgroups (White/Black/Other, male/female), with coefficients for log-transformed age, log-transformed lipids (total cholesterol, HDL-C), log-transformed systolic blood pressure, current smoking, treated hypertension, diabetes, and their interactions with age. For each subgroup, the model calculates a linear predictor as a weighted sum of covariates, then transforms it into a 10-year risk estimate using the baseline survival and mean linear predictor values reported in \cite{goff20142013}:
$$\text{Risk} = 1 - S_0^{\exp(\text{LP} - \overline{\text{LP}})}$$
where $S_0$ is the baseline 10-year survival probability, $\text{LP}$ is the individual's linear predictor, and $\overline{\text{LP}}$ is the mean LP. Individuals classified as ``Other" race are assigned the risk model as White individuals of the same sex, consistent with the original guidelines.

\textit{Covariate distribution:} Patient covariates $X_i$ were generated from the 2017–March 2020 pre-pandemic NHANES public data files. We retained individuals aged 40–79 and selected variables relevant to the Pooled Cohort Equations: age, sex, race/ethnicity, systolic blood pressure, smoking status, diabetes status, blood pressure medication use, HDL cholesterol, and total cholesterol. Missing values were imputed using chained equations with the \textit{mice} package in R, and then completed using a single random draw. Finally, to generate a synthetic cohort, we resampled participants uniformly with replacement from this imputed dataset.

\textit{Outcome distributions:} We considered three outcome variables, each generated from a parametric model of the intervention assignment $A_k$ and baseline risk $\bar{R}_{k,1}$ under the original, unupdated Pooled Cohort Equations. First, attending a prevention program was modeled with a linear probability model:
$$\mathbb{P}(Y_k = 1 \mid \bar{R}_{k,1}, A_k) = \alpha_1 \bar{R}_{k,1} + \alpha_2 A_k \left( \bar{R}_{k,1} +  .5 \right)(1 - \bar{R}_{k,1}),$$
with $\alpha_1 = .10$, and $\alpha_2 = .25 \times 16/9$. This specification results in a maximal treatment effect of $.25$ at a risk level of $.25$.

Second, change in cholesterol from baseline to follow-up (in mg/dL) was modeled as a continuous outcome $Y_k$ with a normal random variable, with conditional mean:
$$\E[Y_k \mid \bar{R}_{m,1}, A_k ] = \beta_1 + \beta_2 A_k \bar{R}_{k,1}$$ 
and  standard deviation $\sigma = 5$. With $\beta_1 = 2$ and $\delta_2 = -10$, the treatment (medication) has no effect on those with no risk ($\bar{R}_{k,1}=0$), but rises to an effect of $-5$ mg/dL for an individual with 50\% risk ($\bar{R}_{k,1}=.5$), translating to a Cohen's $d$ of $1$.

Finally, cardiovascular outcomes were simulated using a deliberately misspecified version of the Pooled Cohort Equations. On the linear predictor scale, we defined
\begin{align*}
\eta_k = \gamma_1 + \gamma_2 \log(-\log(1 - \bar{R}_{k,1})) + \gamma_3 A_k.
\end{align*}
The final event probability was then
$$\mathbb{P}(Y_k = 1 \mid \bar{R}_{k,1}, A_k) = 1 - \exp(-\exp(\eta_k)).$$
The model would be correctly specified under $\gamma_1=\gamma_3 = 0$, $\gamma_2 = 1$, recovering the original PCE. Instead, we intentionally set $\gamma_1 = 0.1, \gamma_2 = 0.9,$ and $\gamma_3 = 0.4$, introducing systematic deviation from the true risk and making model updating a meaningful component of the simulation.

\textit{Threshold updating:} Thresholds for initiating treatment were updated using one of two adaptive strategies, each based on data from all patients who had previously passed through the simulated care pathway. In the first approach, the threshold was recalibrated to achieve a desired treatment rate. For example, to target a 50\% referral rate, the new threshold was set to the median predicted risk among the accumulated cohort, so that approximately half of future patients would be treated. In contrast, the second approach applied our adaptive RD estimator to the observed outcomes and risks from past patients. At each risk level, we computed the implied Cohen's $d$, defined as the estimated treatment effect divided by the pooled standard deviation of outcomes across treatment groups. We then found the risk threshold for which Cohen's $d$ was closest to the value corresponding to the desired NNT. To avoid abrupt shifts, we computed the new threshold as an average between the previously used threshold and this optimal threshold.

\textit{Model updating:} To update the risk prediction model, we implemented two approaches: recalibration and revision. In both cases, updates were based solely on patients who had already completed the simulated care pathway. In the recalibration approach, we re-estimated the coefficients of the Pooled Cohort Equations. A Bernoulli regression with a complementary log–log link was fit, using the original PCE linear predictor and treatment indicator as covariates. This produced new coefficient estimates. Rather than replacing the original model, we took a weighted average of the new estimates and the original coefficients using a weight based on the number $n$ of prior patients: $w = 1/(1 + n/5000)$. When $n = 0$, all weight is placed on the original model; when $n = 5000$, the model is averaged 50/50; as $n \to \infty$, the updated model fully replaces the original.

In the revision approach, we refit a single GLM including all covariates of the Pooled Cohort Equations and a treatment indicator, using a Bernoulli family and complementary log-log link. This refitting was done without stratifying by race or sex, in contrast to the original equations. The newly estimated coefficients were then blended with the original coefficients using the same shrinkage rule as the recalibration approach.

\textit{Estimation:} In our simulations, we used 2 degrees of freedom per treatment group, retained principal components explaining 90\% of the variance, and used a bandwidth $h = .02$. As required by the estimator, we specified GLMs:  logistic regression for binary outcomes, and linear regression for continuous outcomes.

\textit{Comparators:} We compared our adaptive RD estimator to several benchmark methods for estimating the local ATE at the final threshold. As a reference, the true effect was computed from the known data-generating process by first obtaining conditional means and then averaging them using the same kernel-based smoothing and bandwidth $h$ as used in our estimator. A \textit{naive} difference-in-means estimator simply subtracted the average outcomes among treated and untreated individuals, ignoring risk and covariates. A kernel-weighted \textit{outcome regression} estimator fit a GLM of the outcome on treatment and a fixed set of predictors from the Pooled Cohort Equations: age, total cholesterol, HDL cholesterol, systolic blood pressure, treated hypertension, smoking, and diabetes. Treatment effects were estimated via counterfactual predictions, again smoothed using the same kernel and bandwidth. A kernel-weighted \textit{IPW} estimator used these same predictors to estimate the propensity score $\hat{e}(X_j)$ via logistic regression; outcomes were scaled by $A_j/\hat{e}(X_j) - (1 - A_j)/(1 - \hat{e}(X_j))$, and averaged using the same kernel weights. Finally, a kernel-weighted \textit{augmented IPW} estimator combined both the outcome and propensity models via the efficient influence function, which was averaged using the same kernel to yield a doubly robust estimate.

\subsection*{Acknowledgements}
This work was supported through a Patient-Centered Outcomes Research Institute
(PCORI) Project Program Award (ME-2024C1-37689), a seed grant from the University of Wisconsin’s American Family Funding Initiative, and a grant from the Agency for Healthcare Research and Quality (R18HS027735). All statements in this report, including its findings and conclusions, are solely those of
the authors and do not necessarily represent the views of the PCORI, its Board of Governors or Methodology
Committee.


\begin{thebibliography}{10}

\bibitem{challener2019proliferation}
Challener DW, Prokop LJ, Abu-Saleh O.
\newblock The proliferation of reports on clinical scoring systems: issues about uptake and clinical utility.
\newblock JAMA. 2019;321(24):2405-6.

\bibitem{atkins2022developing}
Atkins D, Makridis CA, Alterovitz G, Ramoni R, Clancy C.
\newblock Developing and implementing predictive models in a learning healthcare system: traditional and artificial intelligence approaches in the veterans health administration.
\newblock Annual Review of Biomedical Data Science. 2022;5:393-413.

\bibitem{greene2012implementing}
Greene SM, Reid RJ, Larson EB.
\newblock Implementing the learning health system: from concept to action.
\newblock Annals of internal medicine. 2012;157(3):207-10.

\bibitem{friedman2010achieving}
Friedman CP, Wong AK, Blumenthal D.
\newblock Achieving a nationwide learning health system.
\newblock Science translational medicine. 2010;2(57):57cm29-9.

\bibitem{etheredge2007rapid}
Etheredge LM.
\newblock A rapid-learning health system: what would a rapid-learning health system look like, and how might we get there?
\newblock Health affairs. 2007;26(Suppl1):w107-18.

\bibitem{kelly2019key}
Kelly CJ, Karthikesalingam A, Suleyman M, Corrado G, King D.
\newblock Key challenges for delivering clinical impact with artificial intelligence.
\newblock BMC medicine. 2019;17:1-9.

\bibitem{muralidharan_scoping_2024}
Muralidharan V, Adewale BA, Huang CJ, Nta MT, Ademiju PO, Pathmarajah P, et~al.
\newblock A scoping review of reporting gaps in {FDA}-approved {AI} medical devices.
\newblock npj Digital Medicine. 2024 Oct;7(1):273.
\newblock Publisher: Nature Publishing Group.
\newblock Available from: \url{https://www.nature.com/articles/s41746-024-01270-x}.

\bibitem{moons2015transparent}
Moons KG, Altman DG, Reitsma JB, Ioannidis JP, Macaskill P, Steyerberg EW, et~al.
\newblock Transparent Reporting of a multivariable prediction model for Individual Prognosis or Diagnosis (TRIPOD): explanation and elaboration.
\newblock Annals of internal medicine. 2015;162(1):W1-W73.

\bibitem{collins2015transparent}
Collins GS, Reitsma JB, Altman DG, Moons KG.
\newblock Transparent reporting of a multivariable prediction model for individual prognosis or diagnosis (TRIPOD): the TRIPOD statement.
\newblock Annals of internal medicine. 2015;162(1):55-63.

\bibitem{reilly2006translating}
Reilly BM, Evans AT.
\newblock Translating clinical research into clinical practice: impact of using prediction rules to make decisions.
\newblock Annals of internal medicine. 2006;144(3):201-9.

\bibitem{wolff2019probast}
Wolff RF, Moons KG, Riley RD, Whiting PF, Westwood M, Collins GS, et~al.
\newblock PROBAST: a tool to assess the risk of bias and applicability of prediction model studies.
\newblock Annals of internal medicine. 2019;170(1):51-8.

\bibitem{rubin1980randomization}
Rubin DB.
\newblock Randomization analysis of experimental data: The Fisher randomization test comment.
\newblock Journal of the American statistical association. 1980;75(371):591-3.

\bibitem{rubin1986statistics}
Rubin DB.
\newblock Statistics and causal inference: Comment: Which ifs have causal answers.
\newblock Journal of the American Statistical Association. 1986;81(396):961-2.

\bibitem{motzer1999survival}
Motzer RJ, Mazumdar M, Bacik J, Berg W, Amsterdam A, Ferrara J, et~al.
\newblock Survival and prognostic stratification of 670 patients with advanced renal cell carcinoma.
\newblock Journal of clinical oncology. 1999;17(8):2530-40.

\bibitem{mekhail2005validation}
Mekhail TM, Abou-Jawde RM, BouMerhi G, Malhi S, Wood L, Elson P, et~al.
\newblock Validation and extension of the Memorial Sloan-Kettering prognostic factors model for survival in patients with previously untreated metastatic renal cell carcinoma.
\newblock Journal of clinical oncology. 2005;23(4):832-41.

\bibitem{frank2002outcome}
Frank I, Blute ML, Cheville JC, Lohse CM, Weaver AL, Zincke H.
\newblock An outcome prediction model for patients with clear cell renal cell carcinoma treated with radical nephrectomy based on tumor stage, size, grade and necrosis: the SSIGN score.
\newblock The Journal of urology. 2002;168(6):2395-400.

\bibitem{leibovich2003prediction}
Leibovich BC, Blute ML, Cheville JC, Lohse CM, Frank I, Kwon ED, et~al.
\newblock Prediction of progression after radical nephrectomy for patients with clear cell renal cell carcinoma: a stratification tool for prospective clinical trials.
\newblock Cancer: Interdisciplinary International Journal of the American Cancer Society. 2003;97(7):1663-71.

\bibitem{o2018international}
O’Mahony C, Jichi F, Ommen SR, Christiaans I, Arbustini E, Garcia-Pavia P, et~al.
\newblock International external validation study of the 2014 European Society of Cardiology guidelines on sudden cardiac death prevention in hypertrophic cardiomyopathy (EVIDENCE-HCM).
\newblock Circulation. 2018;137(10):1015-23.

\bibitem{o2014novel}
O'Mahony C, Jichi F, Pavlou M, Monserrat L, Anastasakis A, Rapezzi C, et~al.
\newblock A novel clinical risk prediction model for sudden cardiac death in hypertrophic cardiomyopathy (HCM risk-SCD).
\newblock European heart journal. 2014;35(30):2010-20.

\bibitem{saaristo2005cross}
Saaristo T, Peltonen M, Lindstr{\"o}m J, Saarikoski L, Sundvall J, Eriksson JG, et~al.
\newblock Cross-sectional evaluation of the Finnish Diabetes Risk Score: a tool to identify undetected type 2 diabetes, abnormal glucose tolerance and metabolic syndrome.
\newblock Diabetes and vascular disease research. 2005;2(2):67-72.

\bibitem{lip2010refining}
Lip GY, Nieuwlaat R, Pisters R, Lane DA, Crijns HJ.
\newblock Refining clinical risk stratification for predicting stroke and thromboembolism in atrial fibrillation using a novel risk factor-based approach: the euro heart survey on atrial fibrillation.
\newblock Chest. 2010;137(2):263-72.

\bibitem{kanis2004meta}
Kanis J, Johnell O, De~Laet C, Johansson H, Od{\'e}n A, Delmas P, et~al.
\newblock A meta-analysis of previous fracture and subsequent fracture risk.
\newblock Bone. 2004;35(2):375-82.

\bibitem{kanis2007use}
Kanis J, Od{\'e}n A, Johnell O, Johansson H, De~Laet C, Brown J, et~al.
\newblock The use of clinical risk factors enhances the performance of BMD in the prediction of hip and osteoporotic fractures in men and women.
\newblock Osteoporosis international. 2007;18:1033-46.

\bibitem{kanis2008frax}
Kanis J, Johnell O, Od{\'e}n A, Johansson H, McCloskey E.
\newblock FRAX$^{TM}$ and the assessment of fracture probability in men and women from the UK.
\newblock Osteoporosis international. 2008;19:385-97.

\bibitem{mccarthy2015predictive}
McCarthy JF, Bossarte RM, Katz IR, Thompson C, Kemp J, Hannemann CM, et~al.
\newblock Predictive modeling and concentration of the risk of suicide: implications for preventive interventions in the US Department of Veterans Affairs.
\newblock American journal of public health. 2015;105(9):1935-42.

\bibitem{motzer2010phase}
Motzer RJ, Escudier B, Oudard S, Hutson TE, Porta C, Bracarda S, et~al.
\newblock Phase 3 trial of everolimus for metastatic renal cell carcinoma: final results and analysis of prognostic factors.
\newblock Cancer. 2010;116(18):4256-65.

\bibitem{rini2019atezolizumab}
Rini BI, Powles T, Atkins MB, Escudier B, McDermott DF, Suarez C, et~al.
\newblock Atezolizumab plus bevacizumab versus sunitinib in patients with previously untreated metastatic renal cell carcinoma (IMmotion151): a multicentre, open-label, phase 3, randomised controlled trial.
\newblock The Lancet. 2019;393(10189):2404-15.

\bibitem{ljungberg2015eau}
Ljungberg B, Bensalah K, Canfield S, Dabestani S, Hofmann F, Hora M, et~al.
\newblock EAU guidelines on renal cell carcinoma: 2014 update.
\newblock European urology. 2015;67(5):913-24.

\bibitem{maron2019enhanced}
Maron MS, Rowin EJ, Wessler BS, Mooney PJ, Fatima A, Patel P, et~al.
\newblock Enhanced American College of Cardiology/American Heart Association strategy for prevention of sudden cardiac death in high-risk patients with hypertrophic cardiomyopathy.
\newblock JAMA cardiology. 2019;4(7):644-57.

\bibitem{bartoli2011oral}
Bartoli E, Fra G, Schianca GC.
\newblock The oral glucose tolerance test (OGTT) revisited.
\newblock European journal of internal medicine. 2011;22(1):8-12.

\bibitem{hindricks20212020}
Hindricks G, Potpara T, Dagres N, Arbelo E, Bax JJ, Blomstr{\"o}m-Lundqvist C, et~al.
\newblock 2020 ESC Guidelines for the diagnosis and management of atrial fibrillation developed in collaboration with the European Association for Cardio-Thoracic Surgery (EACTS) The Task Force for the diagnosis and management of atrial fibrillation of the European Society of Cardiology (ESC) Developed with the special contribution of the European Heart Rhythm Association (EHRA) of the ESC.
\newblock European heart journal. 2021;42(5):373-498.

\bibitem{cosman2014clinician}
Cosman F, de~Beur SJ, LeBoff M, Lewiecki E, Tanner B, Randall S, et~al.
\newblock Clinician’s guide to prevention and treatment of osteoporosis.
\newblock Osteoporosis international. 2014;25:2359-81.

\bibitem{reger2019integrating}
Reger GM, McClure ML, Ruskin D, Carter SP, Reger MA.
\newblock Integrating predictive modeling into mental health care: an example in suicide prevention.
\newblock Psychiatric services. 2019;70(1):71-4.

\bibitem{van2023there}
Van~Calster B, Steyerberg EW, Wynants L, van Smeden M.
\newblock There is no such thing as a validated prediction model.
\newblock BMC medicine. 2023;21(1):1-8.

\bibitem{binuya2022methodological}
Binuya M, Engelhardt E, Schats W, Schmidt M, Steyerberg E.
\newblock Methodological guidance for the evaluation and updating of clinical prediction models: a systematic review.
\newblock BMC Medical Research Methodology. 2022;22(1):316.

\bibitem{nakatsugawa2019needs}
Nakatsugawa M, Cheng Z, Kiess A, Choflet A, Bowers M, Utsunomiya K, et~al.
\newblock The needs and benefits of continuous model updates on the accuracy of RT-induced toxicity prediction models within a learning health system.
\newblock International Journal of Radiation Oncology* Biology* Physics. 2019;103(2):460-7.

\bibitem{subbaswamy2020development}
Subbaswamy A, Saria S.
\newblock From development to deployment: dataset shift, causality, and shift-stable models in health AI.
\newblock Biostatistics. 2020;21(2):345-52.

\bibitem{dresser1992wanted}
Dresser R.
\newblock Wanted single{,} white male for medical research.
\newblock The Hastings Center Report. 1992;22(1):24-9.

\bibitem{meltzer_what_2008}
Meltzer LA, Childress JF.
\newblock What {Is} {Fair} {Participant} {Selection?}
\newblock In: Emanuel EJ, Grady CC, Crouch RA, Lie RK, Miller FG, Wendler DD, editors. The {Oxford} {Textbook} of {Clinical} {Research} {Ethics}. Oxford {Textbook} {Ser.}. Oxford University Press; 2008. p. 377-85.

\bibitem{steyerberg2004validation}
Steyerberg EW, Borsboom GJ, van Houwelingen HC, Eijkemans MJ, Habbema JDF.
\newblock Validation and updating of predictive logistic regression models: a study on sample size and shrinkage.
\newblock Statistics in medicine. 2004;23(16):2567-86.

\bibitem{janssen2008updating}
Janssen K, Moons K, Kalkman C, Grobbee D, Vergouwe Y.
\newblock Updating methods improved the performance of a clinical prediction model in new patients.
\newblock Journal of clinical epidemiology. 2008;61(1):76-86.

\bibitem{su2018review}
Su TL, Jaki T, Hickey GL, Buchan I, Sperrin M.
\newblock A review of statistical updating methods for clinical prediction models.
\newblock Statistical methods in medical research. 2018;27(1):185-97.

\bibitem{li2020developing}
Li RC, Asch SM, Shah NH.
\newblock Developing a delivery science for artificial intelligence in healthcare.
\newblock NPJ digital medicine. 2020;3(1):107.

\bibitem{magrabi2019artificial}
Magrabi F, Ammenwerth E, McNair JB, De~Keizer NF, Hypp{\"o}nen H, Nyk{\"a}nen P, et~al.
\newblock Artificial intelligence in clinical decision support: challenges for evaluating AI and practical implications.
\newblock Yearbook of medical informatics. 2019;28(01):128-34.

\bibitem{horwitz2019creating}
Horwitz LI, Kuznetsova M, Jones SA.
\newblock Creating a learning health system through rapid-cycle, randomized testing.
\newblock N Engl J Med. 2019;381(12):1175-9.

\bibitem{heys2023development}
Heys M, Kesler E, Sassoon Y, Wilson E, Fitzgerald F, Gannon H, et~al.
\newblock Development and implementation experience of a learning healthcare system for facility based newborn care in low resource settings: The Neotree.
\newblock Learning Health Systems. 2023;7(1):e10310.

\bibitem{thistlethwaite1960regression}
Thistlethwaite DL, Campbell DT.
\newblock Regression-discontinuity analysis: An alternative to the ex post facto experiment.
\newblock Journal of Educational Psychology. 1960;51(6):309.

\bibitem{hahn2001identification}
Hahn J, Todd P, Van~der Klaauw W.
\newblock Identification and estimation of treatment effects with a regression-discontinuity design.
\newblock Econometrica. 2001;69(1):201-9.

\bibitem{robinson2015evaluation}
Robinson TE, Zhou L, Kerse N, Scott JD, Christiansen JP, Holland K, et~al.
\newblock Evaluation of a New Zealand program to improve transition of care for older high risk adults.
\newblock Australasian journal on ageing. 2015;34(4):269-74.

\bibitem{robinson2017observational}
Robinson T, Jackson R, Wells S, Kerr A, Marshall R.
\newblock An observational study of how clinicians use cardiovascular risk assessment to inform statin prescribing decisions.
\newblock The New Zealand Medical Journal (Online). 2017;130(1463):28-38.

\bibitem{rosenbaum2007interference}
Rosenbaum PR.
\newblock Interference between units in randomized experiments.
\newblock Journal of the american statistical association. 2007;102(477):191-200.

\bibitem{cattaneo2016inference}
Cattaneo MD, Titiunik R, Vazquez-Bare G.
\newblock Inference in regression discontinuity designs under local randomization.
\newblock The Stata Journal. 2016;16(2):331-67.

\bibitem{aronow2017regression}
Aronow PM, Basta NE, Halloran ME.
\newblock The regression discontinuity design under interference: a local randomization-based approach.
\newblock Observational Studies. 2017;3(2):129-33.

\bibitem{goff20142013}
Goff DC, Lloyd-Jones DM, Bennett G, Coady S, D’agostino RB, Gibbons R, et~al.
\newblock 2013 ACC/AHA guideline on the assessment of cardiovascular risk: a report of the American College of Cardiology/American Heart Association Task Force on Practice Guidelines.
\newblock Journal of the American College of Cardiology. 2014;63(25 Part B):2935-59.

\bibitem{richardson2013single}
Richardson TS, Robins JM.
\newblock Single world intervention graphs (SWIGs): A unification of the counterfactual and graphical approaches to causality.
\newblock Center for the Statistics and the Social Sciences, University of Washington Series Working Paper. 2013;128(30):2013.

\bibitem{malinsky2019potential}
Malinsky D, Shpitser I, Richardson T.
\newblock A potential outcomes calculus for identifying conditional path-specific effects.
\newblock In: The 22nd International Conference on Artificial Intelligence and Statistics. PMLR; 2019. p. 3080-8.

\bibitem{stierman2021national}
Stierman B, Afful J, Carroll MD, Chen TC, Davy O, Fink S, et~al.
\newblock National Health and Nutrition Examination Survey 2017-March 2020 prepandemic data files-development of files and prevalence estimates for selected health outcomes.
\newblock National health statistics reports. 2021;(158):10-15620.

\bibitem{furukawa2011obtain}
Furukawa TA, Leucht S.
\newblock How to obtain NNT from Cohen's d: comparison of two methods.
\newblock PloS one. 2011;6(4):e19070.

\bibitem{hudgens2008toward}
Hudgens MG, Halloran ME.
\newblock Toward causal inference with interference.
\newblock Journal of the American Statistical Association. 2008;103(482):832-42.

\bibitem{cattaneo2016interpreting}
Cattaneo MD, Titiunik R, Vazquez-Bare G, Keele L.
\newblock Interpreting regression discontinuity designs with multiple cutoffs.
\newblock The Journal of Politics. 2016;78(4):1229-48.

\bibitem{butler2009regression}
Butler DM.
\newblock A regression discontinuity design analysis of the incumbency advantage and tenure in the US House.
\newblock Electoral Studies. 2009;28(1):123-8.

\bibitem{behaghel2008perverse}
Behaghel L, Cr{\'e}pon B, S{\'e}dillot B.
\newblock The perverse effects of partial employment protection reform: The case of French older workers.
\newblock Journal of Public Economics. 2008;92(3-4):696-721.

\bibitem{patterson2017using}
Patterson BW, Smith MA, Repplinger MD, Pulia MS, Svenson JE, Kim MK, et~al.
\newblock Using chief complaint in addition to diagnosis codes to identify falls in the emergency department.
\newblock Journal of the American Geriatrics Society. 2017;65(9):E135-40.

\bibitem{patterson2018using}
Patterson BW, Repplinger MD, Pulia MS, Batt RJ, Svenson JE, Trinh A, et~al.
\newblock Using the Hendrich II inpatient fall risk screen to predict outpatient falls after emergency department visits.
\newblock Journal of the American Geriatrics Society. 2018;66(4):760-5.

\bibitem{patterson2019training}
Patterson BW, Engstrom CJ, Sah V, Smith MA, Mendon{\c{c}}a EA, Pulia MS, et~al.
\newblock Training and interpreting machine learning algorithms to evaluate fall risk after emergency department visits.
\newblock Medical care. 2019;57(7):560.

\bibitem{jacobsohn2022collaborative}
Jacobsohn GC, Leaf M, Liao F, Maru AP, Engstrom CJ, Salwei ME, et~al.
\newblock Collaborative design and implementation of a clinical decision support system for automated fall-risk identification and referrals in emergency departments.
\newblock In: Healthcare. vol.~10. Elsevier; 2022. p. 100598.

\bibitem{hekman2023effect}
Hekman D, Cochran A, Maru A, Shah M, Liao F, Barton H, et~al.
\newblock Effectiveness of an Emergency Department Based Machine Learning Clinical Decision Support Tool to Prevent Outpatient Falls among Older Adults: A Protocol Paper for a Quasi-experimental Study (Preprint).
\newblock JMIR Research Protocols. 2023 04.

\bibitem{hekman2024dashboarding}
Hekman DJ, Barton HJ, Maru AP, Wills G, Cochran AL, Fritsch C, et~al.
\newblock Dashboarding to Monitor Machine-Learning-Based Clinical Decision Support Interventions.
\newblock Applied Clinical Informatics. 2024;15(01):164-9.

\bibitem{delgado2021unifying}
Delgado C, Baweja M, Crews DC, Eneanya ND, Gadegbeku CA, Inker LA, et~al.
\newblock A unifying approach for GFR estimation: recommendations of the NKF-ASN task force on reassessing the inclusion of race in diagnosing kidney disease.
\newblock Journal of the American Society of Nephrology. 2021;32(12):2994-3015.

\bibitem{lee2008randomized}
Lee DS.
\newblock Randomized experiments from non-random selection in US House elections.
\newblock Journal of Econometrics. 2008;142(2):675-97.

\end{thebibliography}

\end{document}